\documentclass[twocolumn,tighten]{aastex63}
\usepackage{graphicx}
\usepackage{color}
\usepackage{amsmath}
\bibpunct{(}{)}{;}{a}{}{,}
\usepackage[T1]{fontenc} % Use modern font encodings

\DeclareSymbolFont{matha}{OML}{txmi}{m}{it}% txfonts
\DeclareMathSymbol{\varv}{\mathord}{matha}{118}

\shorttitle{Cosmic Ray-Induced $\rm H_2$ Photodissociation}
\shortauthors{Sipil\"a et al.}

\graphicspath{{./}{figures/}}

\begin{document}

\title{New Estimate for the Cosmic Ray-Induced $\rm H_2$ Photodissociation Rate in the Interstellar Medium}

\correspondingauthor{Olli Sipil\"a}
\email{osipila@mpe.mpg.de}
\author{Olli Sipil\"a}
\affiliation{Max-Planck-Institut f\"ur Extraterrestrische Physik, Giessenbachstrasse 1, 85748 Garching, Germany}
\author{Paola Caselli}
\affiliation{Max-Planck-Institut f\"ur Extraterrestrische Physik, Giessenbachstrasse 1, 85748 Garching, Germany}
\author{Marco Padovani}
\affiliation{INAF -- Osservatorio Astrofisico di Arcetri, Largo E. Fermi 5, 50125 Firenze, Italy}
\author{Daniele Galli}
\affiliation{INAF -- Osservatorio Astrofisico di Arcetri, Largo E. Fermi 5, 50125 Firenze, Italy}
\author{Tommaso Grassi}
\affiliation{Max-Planck-Institut f\"ur Extraterrestrische Physik, Giessenbachstrasse 1, 85748 Garching, Germany}
\author{Helgi Rafn Hrodmarsson}
\affiliation{Univ Paris Est Creteil and Universit\'e Paris Cit\'e, CNRS, LISA UMR 7583, F-94010 Cr\'eteil, France}
\author{Sigurd Sigersen Jensen}
\affiliation{Max-Planck-Institut f\"ur Extraterrestrische Physik, Giessenbachstrasse 1, 85748 Garching, Germany}
\author{Evelyne Roueff}
\affiliation{LUX, Observatoire de Paris, PSL Research University, CNRS, Sorbonne Universit\'e, 92190 Meudon, France}

\begin{abstract}
In the interstellar medium, cosmic rays (CRs) generate a field of ultraviolet (UV) photons via the excitation and subsequent radiative decay of $\rm H_2$ molecules. This UV field is a major agent of ionization and dissociation in the inner regions of molecular clouds that are shielded from the effects of the interstellar radiation field. In particular, the dissociation of $\rm H_2$, by far the most abundant molecule in interstellar clouds, leads to the production of atomic hydrogen which then takes part in the production of a multitude of molecules, in particular complex organics on the surfaces of interstellar dust grains. Precise knowledge of the rates of CR-induced dissociation processes is thus crucial for constructing reliable chemical models. For the present paper, we have derived a new value of $k_{\rm diss, CR}(\mbox{$\rm H_2$})=0.831\zeta$ for the rate of $\rm H_2$ dissociation, where $\zeta$ is the CR ionization rate of $\rm H_2$. This prediction contrasts a previous value from the Leiden database which overestimated the rate due to an inconsistent treatment of the $\rm H_2$ abundances and photodissociation cross sections. By running a series of chemical models, we show that the overestimated dissociation rate has a large effect on the results of chemical simulations, with the abundance of methanol being overestimated by over one order of magnitude. Hence, we strongly recommend the adoption of our new estimate $k_{\rm diss, CR}(\mbox{$\rm H_2$})=0.831\zeta$ in all chemical models that include this process. Our newly derived value corresponds to $\rm H_2$ being purely in the para form ($J^{\prime\prime} = 0$). However, in the interiors of molecular clouds the $\rm H_2$ ortho-to-para ratio is low and using the rate for para-$\rm H_2$ is an adequate approximation.
\end{abstract}

\keywords{astrochemistry, cosmic rays, interstellar medium, molecular clouds, molecular processes, photochemistry}

\section{Introduction}

Cosmic rays (CRs) are energetic particles that travel through the interstellar medium at high velocities. They have a large impact on the physical and chemical properties of molecular clouds by providing heating for the gas and serving as the main ionizing agent in dense clouds where ultraviolet (UV) photons from the interstellar radiation field do not penetrate \citep{Herbst73}. CRs passing through dust grains produce a transient high-temperature heating resulting in the desorption of frozen species that sublimate and return to the gas phase \citep{Hasegawa93a}, modifying the composition of interstellar ices and, in a complementary way, the abundance of gas-phase molecules. 

In addition to direct ionization and heating, secondary electrons generated by CR ionization are also responsible for creating a UV field deep within molecular clouds through the excitation and subsequent radiative decay of Lyman-Werner transitions of H$_2$ \citep{Prasad83}. In the gas phase, this CR-induced UV field leads to locally enhanced photoionization and photodissociation of many atomic and molecular species \citep{Sternberg87,Gredel87, Gredel89}; in the dust phase, CR-induced photodissociation of ices is responsible for the formation of radicals that can trigger further chemical reactions on the grain surface, especially in warm regions ($T$ higher than a few tens of K) where evaporation makes otherwise dominant hydrogenation reactions less competitive \citep{Ruffle01, Garrod06}. 

The comprehensive collection of photoionization and photodissociation cross sections made available through the Leiden database\footnote{\tt https://home.strw.leidenuniv.nl/$\sim$ewine/photo/} has recently allowed a careful and systematic re-evaluation of photorates for a variety of interstellar UV fields \citep{Heays17, Hrodmarsson23}. In addition, recent advances on primary CR propagation \citep{Padovani09,Padovani18a,Padovani22}, secondary electron generation \citep{Ivlev21}, as well as the availability of state-of-the-art rovibrationally resolved electron impact excitation cross sections of H$_2$ \citep{Scarlett23} and H$_2$ radiative emission coefficients 
\citep{Abgrall93a,Abgrall93b,Abgrall93c,Abgrall97,Abgrall00,Roueff19}, have made possible a reevaluation of the CR-induced UV spectrum in its line and continuum components in molecular clouds (\citealt{Padovani24}; hereafter P24). 
Photorates computed with the new H$_2$ emission spectrum determined by P24 are, on average, lower by a factor of $\sim 2$ than those computed by \citet{Heays17}, with deviations 
up to a factor of $\sim 5$ for some species, with a few notable exceptions. A particularly striking case is the CR-induced photodissociation rate of H$_2$. No value of the rate is given in the published paper by \citet{Heays17}, but after the 2023 update \citep{Hrodmarsson23} the database reported a value of $k_{\rm diss, CR}({\rm H_2}) = 24.7\zeta$, where $\zeta$ is the CR ionization rate of $\rm H_2$ which was assumed to be $10^{-16} \, \rm s^{-1} \, H_2^{-1}$ for the calculations of the CR ionization rates for all other species in the database. However, it was not clear based on the information in the database, or based on \citet{Heays17}, whether the cross section in the database corresponded to an $\rm H_2$ ortho-to-para ratio of 0:1 or 3:1. Assuming the former, P24 derived a very high value of $k_{\rm diss, CR}({\rm H_2}) = 404\zeta$ using their newly calculated para-$\rm H_2$ luminescence spectrum. However, it has since been deduced that \color{black}the database cross section in fact contains a contribution of ortho-$\rm H_2$, with the $\rm H_2$ ortho-to-para ratio set to 3:1 (H. Hrodmarsson, priv. comm.). Therefore, the predictions of both P24 and the Leiden database are inapplicable to situations where a substantial amount of $\rm H_2$ is in the para form, as is the case in molecular clouds. One in fact expects ortho-to-para ratios ranging from $\sim$0.1 (corresponding to a kinetic temperature of $\sim 40 \, \rm K$) in the transition region from diffuse to molecular clouds, down to $\sim$$10^{-4}$ in the densest regions (see for example the simulations of \citet{Lupi21}). Therefore, applying the value of $24.7\zeta$ to CR-induced $\rm H_2$ photodissociation in molecular cloud conditions is unsound. To prevent any future confusion, the value has now been removed from the database website. 

Precise knowledge of the value of $k_{\rm diss, CR}({\rm H_2})$ is crucial for estimating the efficiencies of chemical reactions that involve atomic hydrogen because $\rm H_2$ is by far the most abundant molecule inside interstellar clouds where its dissociation is the primary source of H atoms. In particular, hydrogenation reactions are thought to be mainly responsible for the formation of many complex molecules (e.g., methanol) on the surfaces of dust grains. Thus, if the $\rm H_2$ dissociation rate is high (low), one expects an increased (decreased) degree of molecular complexity to arise. 

In this paper, we present a new computation of the CR-induced UV photodissociation rate of $\rm H_2$, using the CR-induced emission spectrum computed by P24 applied to a photodissociation cross section that contains the contribution of para-$\rm H_2$ only. We then quantify via chemical simulations the effect of the proposed values for the $k_{\rm diss, CR}({\rm H_2})$ on the abundances of common chemical species in molecular clouds, also checking the impact of the new photorates for molecules other than $\rm H_2$ derived by P24. Finally, we discuss our results and present our conclusions.

\section{CR-Induced $\rm H_2$ Photodissociation} \label{sec:Hydrogen}

We have recalculated the H$_2$, $\varv=0$, $J = 0$ photodissociation rate, integrating
the CR-induced H$_2$ UV spectrum 
computed by P24  in the 75 - 132 nm range over the para-H$_2$ 
cross sections.
This wavelength range encompasses two different mechanisms for H$_2$ photodissociation.

Photons below 118376.99~cm$^{-1}$ or above 84.476~nm (corresponding to H $n=2$ dissociation limit) are not able to directly photodissociate H$_2$ and involve spontaneous radiative dissociation where discrete radiative absorption in the Lyman
($B$-$X$), Werner ($C$-$X$), $B^\prime$-$X$ and $D$-$X$ systems is followed by continuous emission above the dissociation limit of $X$ ground state H$_2$. This range of energies has been extensively studied by \citet{Abgrall93a,Abgrall93b,Abgrall93c, Abgrall97,Abgrall00}.
The wavelength  profile dependent photodissociation cross section from level $\varv^{\prime\prime}=0, J^{\prime\prime}=0$ in the ground electronic state to level $\varv^\prime$, $J^\prime=1$ in the electronic excited state is expressed as
\begin{equation}
\sigma_{\rm diss}(\lambda) = \frac{\pi e^2}{m_e c^2} \lambda^2 \, f_{\rm abs} \,  p_{\varv^\prime, 1}^{\rm diss}\,\phi(\lambda),
\end{equation}
where  $p_{\varv^\prime, 1}$ is the dissociation probability of the upper
$\varv^\prime$, $J^\prime=1 $ electronically excited level, and $\phi(\lambda)$ is the normalized profile of the absorbing transition such that $\int {\phi(\lambda) d\lambda = 1 }$. Assuming that the wavelength variation over the profile is small, one can derive the wavelength-integrated photodissociation cross section of a specific transition as below:
\begin{equation}
\sigma_{\rm diss}\equiv \int{\sigma_{\rm diss}(\lambda) d\lambda} = 10^{-7}\frac{\pi e^2}{m_e c^2} \lambda^2 \, f_{\rm abs} \,  p_{\varv^\prime, 1}^{\rm diss}\,    ,
\end{equation}
where the $10^{-7}$ factor arises if the wavelength is expressed in nm and $\sigma_{\rm diss}$ in cm$^2$~nm, leading to 
\begin{equation}
\sigma_{\rm diss} = 8.8528\times 10^{-20} \, \lambda^2 \, f_{\rm abs}\, \, p_{\varv^\prime, 1}^{\rm diss}~{\rm cm}^2~{\rm nm}.
\end{equation}

The different parameters are displayed in Table \ref{tab:photodissociation1}
for the  $R(0)$ transitions of para-H$_2$:   
$E-X$ specifies the  electronic band system involved where $E$ is the upper electronic state and $X$ is the H$_2$ ground electronic state. $f_{\rm abs}$ is the absorption line oscillator strength and 
$p_{\varv^\prime, 1}^{\rm diss}\,$ is the dissociation probability of the upper $E$, $\varv^\prime, J'=1$
level, that varies significantly depending on $\varv^\prime$ and $E$. We also display $\bar\nu$, the transition wavenumber, $A_{\rm tot}(\varv^\prime, 1)$, the total decaying probability of the upper level that contributes to the Lorentzian profile of the dissociating transitions.
The profile results from the convolution of the Lorentzian profile whose half-width at half maximum is proportional to the inverse of the total decay rate ($A_{\rm tot}$) with the Gaussian profile corresponding to Doppler broadening (with a $b$ parameter assumed here of 1~km~s$^{-1}$) due to thermal and turbulent motions.
The photodissociation rate, expressed in s$^{-1}$,  involves the sum over all dissociating transitions of the product 
\begin{equation}
    k_{\rm diss} = \int{ \sigma_{\rm diss}(\lambda) I(\lambda) d\lambda} ,
\end{equation}
 where $I(\lambda)$ is the radiation field intensity expressed in photons~cm$^{-2}$~s$^{-1}$~nm$^{-1}$.
The  75 - 84.476~nm lower wavelength interval involves photon energies above the $n=2$ dissociation limit that lead to photodissociation through levels that radiatively decay 
towards the ground state and at the same time can dissociate through 
non adiabatic couplings resulting in a predissociative broadening. Detailed theoretical
and experimental studies have been performed in the groups of M. Glass-Maujean and H. Schmoranzer
\citep{Glass-Maujean2012a,Glass-Maujean2012b,Glass-Maujean2013} and the corresponding data are reported on the MOLAT website\footnote{https://molat.obspm.fr}.

\begin{table*}
\caption{Wavelength-integrated H$_2$ photodissociation cross sections
due to $E - X , \, \varv^\prime - 0,\,  R(0)$ transitions above 84.47~nm. Numbers in parentheses are exponential. 
\label{tab:photodissociation1}}
%\centering%
\begin{tabular}{ll|cccccc}
\hline
& transition &    $\lambda$ & $\bar{\nu}$ & $f_{\rm abs}$  &   $p_{\varv^\prime, 1}^{\rm diss}$  & $A_{\rm tot}(\varv^\prime, 1)$ & $\sigma_{\rm diss}$  \\
$\varv^\prime$ & $E$-$X$   &    (nm)     &    (cm$^{-1}$) &   &       &  (s$^{-1}$) & (cm$^2$~nm) \\
\hline
  0 &      $B$-$X$    &   110.8127    &    90242.364     &   1.665(-3)    &    4.309(-9)    &    1.863(9)    &    7.799(-27) \\
  1 &       $B$-$X$    &   109.2193    &    91558.909     &   5.783(-3)    &    7.680(-8)    &    1.738(9)    &    4.690(-25) \\
  2 &       $B$-$X$    &   107.7136    &    92838.787     &   1.167(-2)    &    1.335(-5)    &    1.630(9)    &    1.600(-22) \\
  3 &      $B$-$X$    &   106.2879    &    94084.087     &   1.789(-2)    &    1.541(-4)    &    1.534(9)    &    2.757(-21) \\
  4 &       $B$-$X$    &   104.9364    &    95295.817     &   2.319(-2)    &    1.280(-3)    &    1.448(9)    &    2.894(-20) \\
  5 &      $B$-$X$    &   103.6543    &    96474.531     &   2.683(-2)    &    1.862(-2)    &    1.371(9)    &    4.752(-19) \\
  6 &       $B$-$X$    &   102.4370    &    97620.977     &   2.871(-2)    &    3.334(-2)    &    1.301(9)    &    8.892(-19) \\
  7 &      $B$-$X$    &   101.2810    &    98735.202     &   2.970(-2)    &    2.052(-1)    &    1.236(9)    &    5.534(-18) \\
  0 &       $C$-$X$    &   100.8550    &    99152.248     &   4.395(-2)    &    1.382(-4)    &    1.180(9)    &    5.469(-21) \\
  8 &       $B$-$X$    &   100.1821    &    99818.231     &   2.677(-2)    &    3.123(-1)    &    1.177(9)    &    7.428(-18) \\
  9 &       $B$-$X$    &    99.1376    &   100869.882     &   2.612(-2)    &    4.126(-1)    &    1.122(9)    &    9.377(-18) \\
  1 &       $C$-$X$    &    98.5630    &   101457.971     &   6.890(-2)    &    3.982(-4)    &    1.159(9)    &    2.360(-20) \\
 10 &      $B$-$X$    &    98.1436    &   101891.473     &   2.069(-2)    &    4.101(-1)    &    1.072(9)    &    7.235(-18) \\
 11 &      $B$-$X$    &    97.1984    &   102882.331     &   2.004(-2)    &    4.072(-1)    &    1.025(9)    &    6.825(-18) \\
  2 &      $C$-$X$    &    96.4979    &   103629.155     &   6.860(-2)    &    1.633(-3)    &    1.138(9)    &    9.235(-20) \\
 12 &      $B$-$X$    &    96.2976    &   103844.705     &   1.316(-2)    &    5.200(-1)    &    9.815(8)    &    5.618(-18) \\
 13 &       $B$-$X$    &    95.4412    &   104776.565     &   1.416(-2)    &    5.097(-1)    &    9.402(8)    &    5.820(-18) \\
  3 &       $C$-$X$    &    94.6422    &   105661.122     &   6.192(-2)    &    1.418(-1)    &    1.052(9)    &    6.962(-18) \\
 14 &      $B$-$X$    &    94.6168    &   105689.420     &   1.340(-3)    &    3.454(-1)    &    9.714(8)    &    3.668(-19) \\
 15 &   $B$-$X$    &    93.8467    &   106556.757     &   9.510(-3)    &    5.597(-1)    &    8.672(8)    &    4.150(-18) \\
 16 &      $B$-$X$    &    93.1062    &   107404.256     &   1.067(-2)    &    5.507(-1)    &    8.356(8)    &    4.509(-18) \\
  4 &    $C$-$X$    &    92.9530    &   107581.262     &   3.419(-2)    &    2.931(-3)    &    1.103(9)    &    7.665(-20) \\
 17 &    $B$-$X$    &    92.3984    &   108226.994     &   6.167(-3)    &    6.064(-1)    &    8.038(8)    &    2.826(-18) \\
 18 &    $B$-$X$    &    91.7251    &   109021.377     &   6.175(-3)    &    6.089(-1)    &    7.755(8)    &    2.801(-18) \\
  5 &      $C$-$X$    &    91.4395    &   109361.976     &   2.424(-2)    &    1.367(-3)    &    1.090(9)    &    2.453(-20) \\
 19 &      $B$-$X$    &    91.0820    &   109791.153     &   3.865(-3)    &    6.347(-1)    &    7.490(8)    &    1.802(-18) \\
  0 &     $B^\prime$-$X$   &    90.4736    &   110529.440     &   2.372(-3)    &    1.833(-4)    &    3.809(8)    &    3.151(-22) \\
 20 &     $B$-$X$    &    90.4701    &   110533.805     &   3.887(-3)    &    6.375(-1)    &    7.242(8)    &    1.795(-18) \\
  6 &      $C$-$X$    &    90.0797    &   111012.828     &   1.645(-2)    &    2.630(-3)    &    1.077(9)    &    3.108(-20) \\
 21 &      $B$-$X$    &    89.8855    &   111252.699     &   2.246(-3)    &    6.392(-1)    &    7.026(8)    &    1.027(-18) \\
 22 &       $B$-$X$    &    89.3298    &   111944.751     &   2.473(-3)    &    6.523(-1)    &    6.805(8)    &    1.140(-18) \\
  1 &      $B^\prime$-$X$   &    88.9643    &   112404.592     &   6.988(-3)    &    7.132(-7)    &    3.236(8)    &    3.492(-24) \\
  7 &     $C$-$X$    &    88.8649    &   112530.356     &   1.149(-2)    &    1.194(-2)    &    1.059(9)    &    9.591(-20) \\
 23 &     $B$-$X$    &    88.7982    &   112614.946     &   7.844(-4)    &    6.341(-1)    &    6.716(8)    &    3.472(-19) \\
  0 &     $D$-$X$    &    88.5462    &   112935.444     &   7.839(-3)    &    1.027(-8)    &    3.549(8)    &    5.588(-26) \\
 24 &    $B$-$X$    &    88.2954    &   113256.145     &   1.601(-3)    &    6.797(-1)    &    6.444(8)    &    7.510(-19) \\
 25 &     $B$-$X$    &    87.8186    &   113871.054     &   3.284(-3)    &    6.149(-1)    &    6.612(8)    &    1.379(-18) \\
  8 &      $C$-$X$    &    87.7824    &   113918.039     &   4.755(-3)    &    6.928(-2)    &    1.032(9)    &    2.247(-19) \\
  2 &     $B^\prime$-$X$   &    87.6284    &   114118.238     &   8.154(-3)    &    6.892(-7)    &    2.658(8)    &    3.820(-24) \\
 \hline
\end{tabular}
 \end{table*} 
 
 \begin{table*}
\caption{Continued from Table~\ref{tab:photodissociation1}.}
%\centering%
\begin{tabular}{ll|cccccc}
\hline
& transition &    $\lambda$ & $\bar{\nu}$ & $f_{\rm abs}$  &   $p_{\varv^\prime, 1}^{\rm diss}$  & $A_{\rm tot}(\varv^\prime, 1)$ & $\sigma_{\rm diss}$  \\
$\varv^\prime$ & $E$-$X$   &    (nm)     &    (cm$^{-1}$) &   &       &  (s$^{-1}$) & (cm$^2$~nm) \\
\hline
26 &     $B$-$X$    &    87.3607    &   114467.921     &   1.047(-3)    &    7.161(-1)    &    6.164(8)    &    5.066(-19) \\
 27 &    $B$-$X$    &    86.9306    &   115034.239     &   1.293(-3)    &    7.229(-1)    &    6.107(8)    &    6.253(-19) \\
  9 &    $C$-$X$    &    86.8402    &   115153.989     &   3.612(-3)    &    6.669(-2)    &    1.060(9)    &    1.608(-19) \\
  1 &    $D$-$X$    &    86.8385    &   115156.295     &   1.517(-2)    &    3.159(-4)    &    3.545(8)    &    3.199(-21) \\
 28 &   $B$-$X$    &    86.5230    &   115576.165     &   6.869(-4)    &    7.521(-1)    &    6.027(8)    &    3.424(-19) \\
  3 &    $B^\prime$-$X$   &    86.4699    &   115647.207     &   6.675(-3)    &    4.515(-6)    &    2.094(8)    &    1.995(-23) \\
 29 &    $B$-$X$    &    86.1418    &   116087.633     &   7.873(-4)    &    7.686(-1)    &    6.073(8)    &    3.975(-19) \\
 10 &    $C$-$X$    &    86.0308    &   116237.414     &   2.682(-3)    &    3.862(-1)    &    1.067(9)    &    6.787(-19) \\
 30 &     $B$-$X$    &    85.7855    &   116569.776     &   4.577(-4)    &    7.919(-1)    &    6.133(8)    &    2.361(-19) \\
  4 &     $B^\prime$-$X$   &    85.5156    &   116937.686     &   7.588(-3)    &    1.400(-5)    &    1.583(8)    &    6.877(-23) \\
 31 &     $B$-$X$    &    85.4589    &   117015.286     &   5.190(-4)    &    8.109(-1)    &    6.265(8)    &    2.721(-19) \\
 11 &  $C$-$X$    &    85.3602    &   117150.616     &   1.723(-3)    &    4.269(-1)    &    1.082(9)    &    4.745(-19) \\
  2 &    $D$-$X$    &    85.2863    &   117252.071     &   1.324(-2)    &    6.838(-7)    &    3.520(8)    &    5.830(-24) \\
 32 &      $B$-$X$    &    85.1636    &   117421.044     &   2.986(-4)    &    8.398(-1)    &    6.440(8)    &    1.610(-19) \\
 33 &      $B$-$X$    &    84.9079    &   117774.685     &   3.262(-4)    &    8.685(-1)    &    6.744(8)    &    1.808(-19) \\
 12 &      $C$-$X$    &    84.8402    &   117868.721     &   1.108(-3)    &    6.309(-1)    &    1.108(9)    &    4.454(-19) \\
  5 &      $B^\prime$-$X$   &    84.8293    &   117883.785     &   3.007(-3)    &    2.581(-3)    &    1.345(8)    &    4.944(-21) \\
 34 &     $B$-$X$    &    84.7009    &   118062.458     &   1.619(-4)    &    9.080(-1)    &    7.212(8)    &    9.337(-20) \\
  6 &     $B^\prime$-$X$   &    84.6183    &   118177.715     &   3.707(-4)    &    3.796(-1)    &    3.435(8)    &    8.920(-20) \\
 35 &     $B$-$X$    &    84.5613    &   118257.406     &   1.137(-4)    &    9.532(-1)    &    8.001(8)    &    6.861(-20) \\
  7 &     $B^\prime$-$X$   &    84.5465    &   118278.121     &   3.270(-4)    &    9.157(-1)    &    3.927(8)    &    1.895(-19) \\
 36 &      $B$-$X$    &    84.5034    &   118338.476     &   1.776(-4)    &    9.816(-1)    &    8.191(8)    &    1.102(-19) \\
  8 &     $B^\prime$-$X$   &    84.5011    &   118341.699     &   6.057(-5)    &    9.610(-1)    &    5.050(8)    &    3.679(-20) \\
 13 &      $C$-$X$    &    84.4952    &   118349.918     &   5.353(-4)    &    8.141(-1)    &    1.126(9)    &    2.754(-19) \\
 37 &    $B$-$X$    &    84.4846    &   118364.711     &   5.586(-7)    &    9.939(-1)    &    1.005(9)    &    3.508(-22) \\
  9 &      $B^\prime$-$X$   &    84.4762    &   118376.593     &   9.314(-5)    &    9.600(-1)    &    1.099(8)    &    5.649(-20) \\
 \hline
\end{tabular}
 \end{table*} 

We finally obtain $k_{\rm diss, CR}(\mbox{p$\rm H_2$})=0.831\zeta$, which is lower than the value previously assigned to $\rm H_2$ ($24.7\zeta$) found in the database tables, but much closer to the assumed primary CR ionization rate of $\rm H_2$ (i.e., 1.000$\zeta$) which was used for computing the CR ionization rates for all other species in the database \citep{Heays17,Hrodmarsson23}. The inconsistent treatment of $\rm H_2$ that gave rise to the $24.7\zeta$ value did not impact computed rates of any other species in the database.

We emphasize that the contribution of the short wavelength range below $83.9$~nm 
contributes to about 50\%. We only consider para-$\mathrm{H_2}$ here as the corresponding ortho/para ratio is expected to be very small in molecular cores such as L1544 that is considered below \citep{Walmsley2004,Furuya16,Lupi21}. Finally, we note that the newly derived value ($0.831\zeta$) for the CR-induced $\rm H_2$ photodissociation rate is of the same order of magnitude as the rate of collisional CR-induced ${\rm H_2}$ dissociation due to secondary electrons, as calculated by \citet{Padovani18b}. The collisional process is distinct from the photodissociation due to CR-generated UV field, which is the focus of the present paper.

\section{Chemical modeling}\label{s:simulations}

We have modeled the chemical evolution in four different scenarios, whose details are collected in Table~\ref{tab:simulationParameters}, to explore the effect of CR-induced photochemistry in molecular clouds. The first one, our fiducial simulation (S1), uses a chemical network (including deuterium and spin-state chemistry; see also below) based on the 2014 public release of the KIDA network \citet{Wakelam15} that we have used in many previous works (e.g., \citealt{Sipila19a,Caselli22}). Notably, the KIDA network does not include the CR-induced photodissociation of $\rm H_2$ at all. For our previous work, we have updated some rate coefficients for CR-induced photodissociation processes from Table~20 in \citet{Heays17}, but -- as already noted in the Introduction -- there the $\rm H_2$ dissociation and ionization rate coefficients are listed as "--" and hence the CR-induced $\rm H_2$ dissociation process was not included in our models either. In the second simulation (S2), we modify the chemical network used in S1 by including the updated photodissociation and photoionization rate coefficients from P24 -- except for $\rm H_2$ dissociation (we get back to this point in Discussion and Conclusions). The remaining two simulations (S3 and S4) then add the $\rm H_2$ photodissociation process with different values for the dissociation rate coefficient: $0.831\zeta$ derived here (S3), or $24.7\zeta$ (S4).

\begin{figure}
\centering
        \includegraphics[width=\columnwidth]{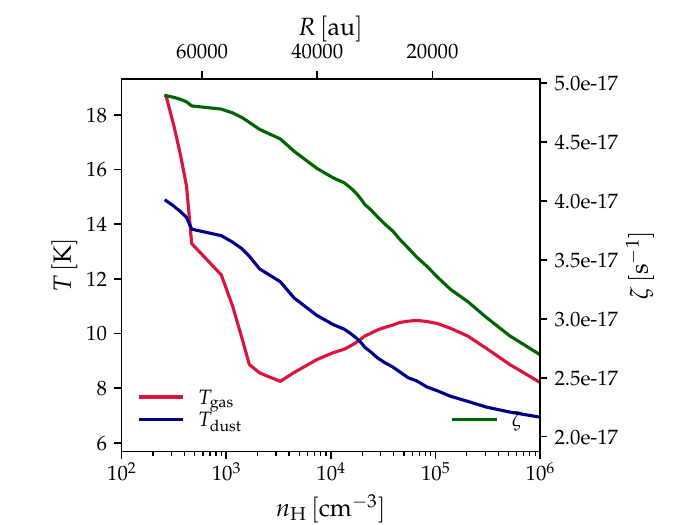}
    \caption{Gas and dust temperatures (scale on left-hand $y$-axis) and $\zeta$ (right-hand $y$-axis) as a function of gas density (bottom $x$-axis)  expressed as the total hydrogen nuclei number density $n_{\rm H} = 2n({\rm H_2}) + n({\rm H})$ or distance from the center of the core in astronomical units (top $x$-axis) in the L1544 model.}
        \label{fig:physicalModel}
\end{figure}

To facilitate probing the effect of the updated rate coefficients of the CR-induced photodissociation reactions on the abundances of common molecules in varying physical conditions, we have adopted the one-dimensional model for the prestellar core L1544 \citep{Keto15} as a template physical model. This allows us to track the changes in gas density and (gas and dust) temperature as a function of location. Similar to \citet{Sipila19a}, the L1544 model is divided into concentric spherical shells, and the chemical simulations have been run separately in each shell to produce abundance radial profiles as a function of time. We have calculated a radial profile for $\zeta$ using the ``Low'' model of \citet{Padovani18a}. Their parametrized formula for $\zeta$ is expressed as a function of $\rm H_2$ column density; we have derived the column density profile by calculating the visual extinction ($A_{\rm V}$) assuming 1\,mag of extinction outside the core, and then converting the extinction in each point in the core into column density according to $N({\rm H_2})/A_{\rm V} = 1.2\times 10^{21}\,\rm cm^{-2} \, mag^{-1}$. The resulting $\zeta$ profile, as well as the gas and dust temperature profiles from \citet{Keto15}, as a function of gas density are shown in Fig.\,\ref{fig:physicalModel}. We stress that we employ the L1544 core model purely for the purposes of having a self-consistently derived representation of how the physical conditions vary with density in dense clouds -- our results are not meant to be interpreted as being specific to L1544.

\begin{table}
\begin{center}
\caption{Set of simulations considered in this paper.}
\begin{tabular}{ll}
\hline
Label & Details \\
\hline
S1 &  Fiducial model not including the rate \\
  & coefficient updates from P24\\
S2 &  As S1, but including updated rate \\
  & coefficients from P24 and neglecting\\
  & CR-induced $\rm H_2$ photodissociation\\
S3 & As S2, but including CR-induced $\rm H_2$ \\
 &  photodissociation with $k_{\rm diss, CR}({\rm H_2}) = 0.831 \zeta$\\
S4 & As S3, but with $k_{\rm diss, CR}({\rm H_2}) = 24.7 \zeta$\\
\hline
\label{tab:simulationParameters}
\end{tabular}
\end{center}
\end{table}

The chemical simulations have been run using our chemical code {\sl pyRate}. In brief, {\sl pyRate} is a gas-grain chemical model that solves a network of rate equations expressed as ordinary differential equations, connecting the gas-phase and grain-surface chemical evolution via adsorption and several desorption mechanisms. The basic functionality of the code is described in \citet{Sipila15a}. We leave further details out of the present paper for brevity. The chemical networks employed here include deuterium and spin-state chemistry, implemented according to the proton hop scenario \citep{Sipila19b}; we have recently found that the proton hop scenario describes some chemical systems better than the full scrambling model does \citep{Harju24,Harju25,JimenezRedondo25}. In the present paper, however, we focus solely on hydrogenated species and do not show results for deuterated ones. The initial abundances, which are the same throughout the model cloud, are displayed in Table~\ref{tab:initialabundances}. We note that our chemical model tracks the $\rm H_2$ ortho-to-para ratio time-dependently and hence an initial value for the ratio is required, but because of the current lack of separate rate coefficients for the CR-induced photodissociation of ortho and para $\rm H_2$, in each simulation we assume $k_{\rm diss, CR}(\mbox{o$\rm H_2$}) = k_{\rm diss, CR}(\mbox{p$\rm H_2$}) = k_{\rm diss, CR}(\mbox{$\rm H_2$})$.

\begin{table}
        \centering
        \caption{Initial abundances (with respect to $n_{\rm H}$). The initial $\rm H_2$ ortho/para ratio is $10^{-3}$ corresponding to thermal equilibrium at 
        the edge of the cloud.}
        \begin{tabular}{ll}
                \hline
                Species & Abundance\\
                \hline
                $\rm H_2$ & $5.00\times10^{-1}$\\
                $\rm He$ & $9.00\times10^{-2}$\\
                $\rm C^+$ & $1.20\times10^{-4}$\\
                $\rm N$ & $7.60\times10^{-5}$\\
                $\rm O$ & $2.56\times10^{-4}$\\
                $\rm S^+$ & $8.00\times10^{-8}$\\
                $\rm Si^+$ & $8.00\times10^{-9}$\\
                $\rm Na^+$ & $2.00\times10^{-9}$\\
                $\rm Mg^+$ & $7.00\times10^{-9}$\\
                $\rm Fe^+$ & $3.00\times10^{-9}$\\
                $\rm P^+$ & $2.00\times10^{-10}$\\
                $\rm Cl^+$ & $1.00\times10^{-9}$\\
                $\rm HD$ & $1.60\times10^{-5}$\\
                \hline
        \end{tabular}
        \label{tab:initialabundances}
\end{table}

\section{Results}
\label{s.results}

\begin{figure*}
\centering
        \includegraphics[width=2\columnwidth]{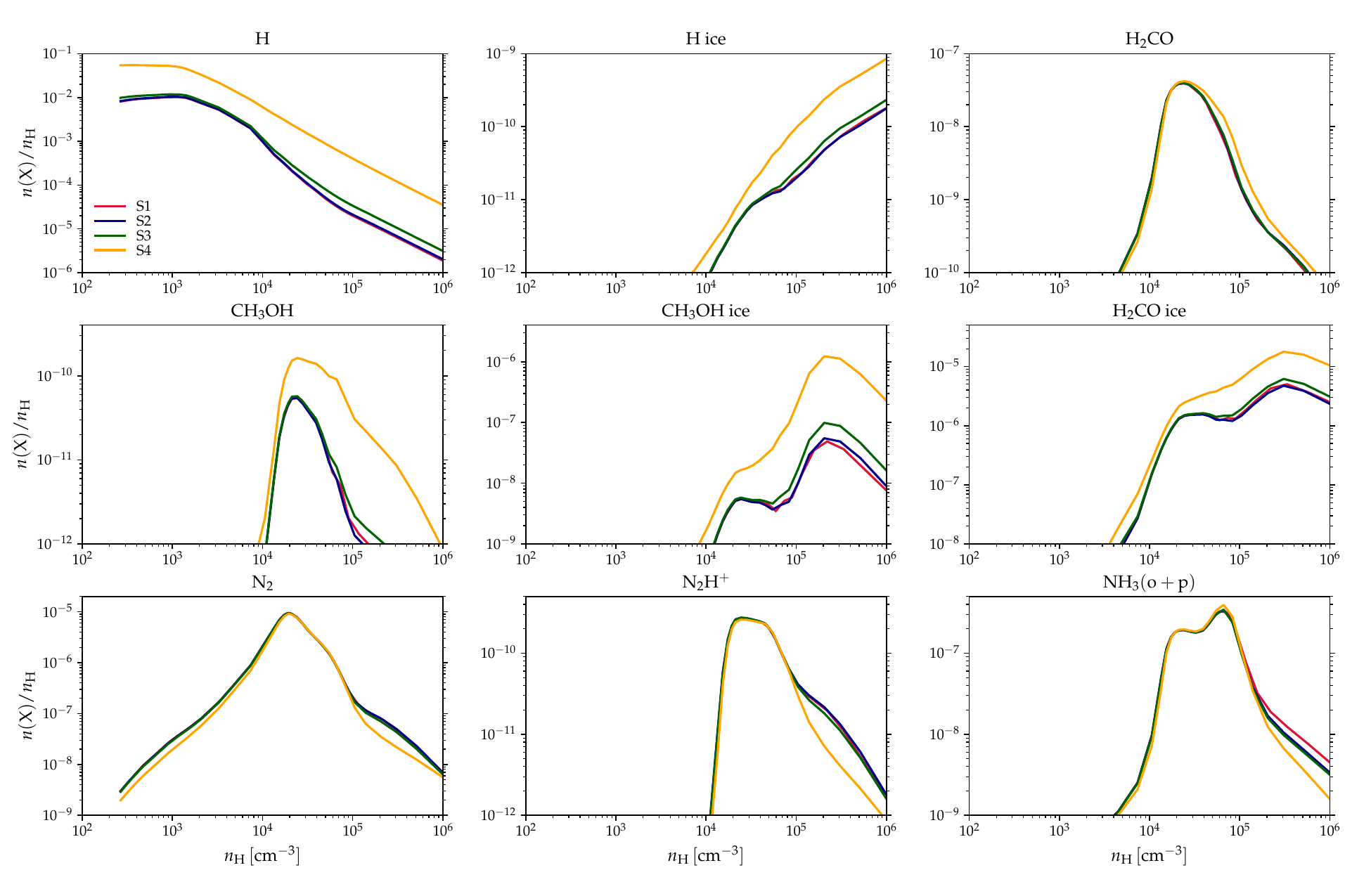}
    \caption{Abundances of selected species, labeled on top of each panel, as a function of gas density, expressed as the total hydrogen nuclei number density $n_{\rm H} = 2n({\rm H_2}) + n({\rm H})$. The four curves correspond to the four different simulation cases, as indicated in the legend in the upper left-hand panel. The denomination ``X~ice'' refers to species X on grain surfaces.}
        \label{fig:abundances}
\end{figure*}

Figure~\ref{fig:abundances} shows the abundance profiles of selected molecules as predicted in the four simulation cases. Because the physical model is static, the evolutionary time is a free parameter; the results are plotted for $t = 10^6$\,yr. We have, however, checked that the trends in the results that we discuss in what follows are qualitatively similar at earlier times as well, for example $10^5$\,yr.

Let us first compare the results of the fiducial simulation (S1) and the updated model without CR-induced $\rm H_2$ photodissociation added (S2). The simulation results are very similar in the two cases. This means that the rate coefficient updates from P24 (excluding $\rm H_2$) have only a limited overall impact on the chemistry -- at least for the species plotted here and for the present physical conditions. However, $\rm N_2$ deserves closer examination, as it is one of the few species for which P24 found larger (a factor of a few level) changes in the dissociation rate coefficients compared to earlier literature values. As noted in the Simulations section, our fiducial model adopts CR-induced photodissociation rates from Table~20 in \citet{Heays17}, and the data given in that table assumes significant self-shielding for $\rm N_2$. The shielded rate coefficient ($39 \zeta$) happens to be close to the new value derived by P24 ($34 \zeta$ for $\rm H_2$ o:p = 0:1 and $R_{\rm V} = 3.1$). If we had used the unshielded rate coefficient value, the difference in the predictions of \citet{Heays17} and P24 on the $\rm N_2$ dissociation rate would be a factor of a few (see also discussion in Sect.\,6 in P24) and the $\rm N_2$ abundances predicted here by the S1 and S2 simulations would be further apart.

The S3 simulation introduces the CR-induced $\rm H_2$ photodissociation process with the newly derived rate coefficient of $0.831 \zeta$. The inclusion of this process does affect the results, but only to a rather limited degree. In particular, the amount of gas-phase atomic H increases with respect to simulations S1 and S2, and this translates to increased amounts of formaldehyde and methanol production on grains via the adsorption and subsequent reactivity of atomic~H. The abundance of gas-phase methanol increases accordingly due to chemical desorption of grain-surface methanol. For gas-phase formaldehyde, the effect is smaller, which is likely due to the fact that, unlike methanol, formaldehyde can also be produced efficiently in the gas phase.

\begin{figure*}
\centering
        \includegraphics[width=2.0\columnwidth]{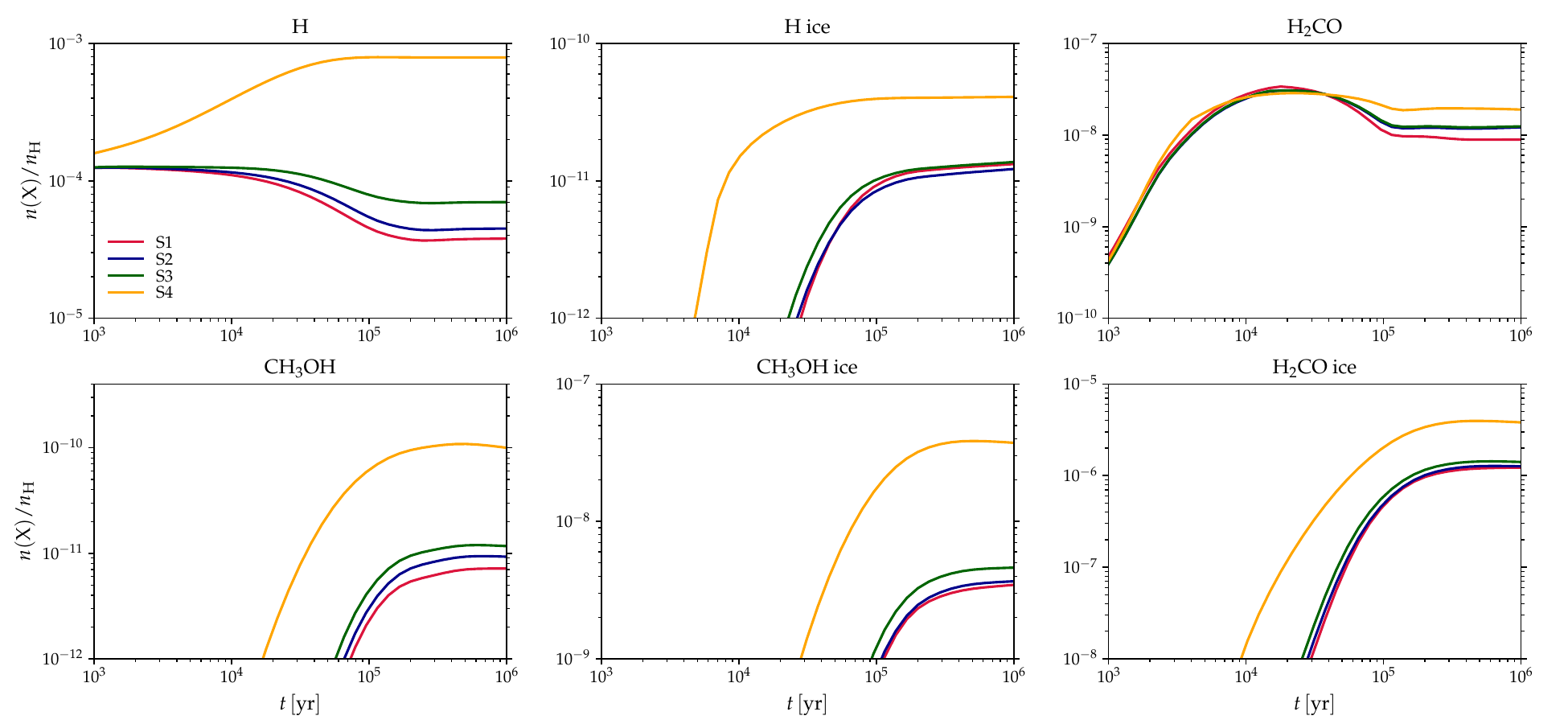}
    \caption{Abundances of selected species, labeled on top of each panel, as a function of time in the model cell corresponding to $n({\rm H}) \sim 3 \times 10^4 \, \rm cm^{-3}$. The four curves correspond to the four different simulation cases, as indicated in the legend in the upper left-hand panel. The denomination ``X~ice'' refers to species X on grain surfaces.}
        \label{fig:timeDependence}
\end{figure*}

Increasing $k_{\rm diss, CR}(\mbox{$\rm H_2$})$ to $24.7 \zeta$ (S4) leads to larger amounts of atomic H in the gas phase, which translates to greatly increased production rates of grain-surface formaldehyde and methanol. The difference in the predicted gas-phase methanol abundance between simulations S3 and S4 is over one order of magnitude at the highest densities. Conversely, simulation S4 predicts a lower amount of gas-phase $\rm N_2$ and $\rm N_2H^+$ at high densities, compared to S3. This is because of increased production of ammonia on grains (again due to higher flux of adsorbing atomic H), which locks a significant part of the nitrogen reservoir in the ice. In the case of ammonia, it is noteworthy that its formation on grains through H + $\rm NH_2$ can lead to chemical desorption of $\rm NH_3$, analogously to methanol which can be chemically desorbed when formed via H + $\rm CH_3O/CH_2OH$. However, the formation and destruction of gas-phase ammonia are governed by gas-phase reactions, while the origin of gas-phase methanol is almost exclusively chemical desorption. This is why gas-phase ammonia does not show a strong increase with the increase of the $\rm H_2$ photodissociation rate, while methanol does.

One general trend common to all four simulations is that the abundance of grain-surface methanol decreases toward the highest densities. This is an artifact caused by the use of a static physical model; the central regions of the model core are very cold ($T < 8$\,K; see Fig.\,\ref{fig:physicalModel}), which prevents even hydrogen from diffusing efficiently around the grain surface, inhibiting the formation of methanol (we reiterate that the same, mostly atomic, initial conditions are used across the model core). If we were to use a time-dependent collapse model, we would obtain high grain-surface methanol abundances also in the central regions (see, e.g., Fig.\,4 in \citealt{Sipila24}).

To complement Fig.\,\ref{fig:abundances}, we show in Fig.\,\ref{fig:timeDependence} the time-evolution of selected species included in the methanol formation chain. The figure confirms that the trends evident in Fig.\,\ref{fig:abundances} are not simply a result of the particular choice of evolutionary time at which the radial profiles are extracted from the simulation data -- the predictions of the S4 model are different to the other models already from early stages of chemical evolution.

\begin{figure}
\centering
        \includegraphics[width=1.0\columnwidth]{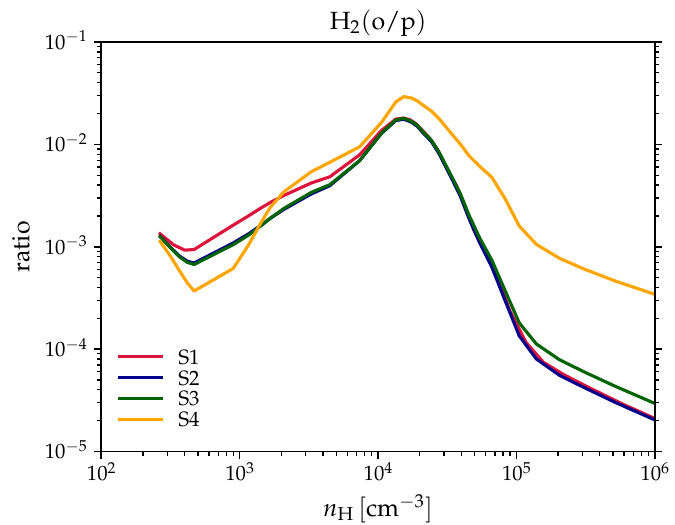}
    \caption{$\rm H_2$ ortho/para (o/p) ratio as a function of gas density, expressed as the total hydrogen nuclei number density $n_{\rm H} = 2n({\rm H_2}) + n({\rm H})$. The four curves correspond to the four different simulation cases, as indicated in the legend.}
        \label{fig:H2_op}
\end{figure}

Figure~\ref{fig:H2_op} shows the $\rm H_2$ ortho/para ratio as predicted by the four simulations. Firstly, we highlight that the use of $24.7\zeta$ for the rate of $\rm H_2$ dissociation leads to a drastic overestimation of the ortho/para ratio by an order of magnitude in the inner core. This occurs because the fast $\rm H_2$ dissociation leads to an increase of atomic H which ends up on grains (cf. Fig.\,\ref{fig:abundances}); $\rm H_2$ forms on grains in an ortho/para ratio of 3, and the subsequent desorption of the ortho-rich $\rm H_2$ ice component then boosts the gas-phase ratio. Our results show clearly that assuming an ortho/para ratio of 3:1 when deriving the $\rm H_2$ photodissociation rate is, under molecular cloud conditions, completely inappropriate. Secondly, the $\rm H_2$ ortho/para ratio is overall higher at intermediate densities compared to the outer core. This is because at low densities the $\rm H_2$ ortho-to-para conversion proceeds mainly via reactions of $\rm H_2$ with $\rm H^+$ with little contribution from grain-surface $\rm H_2$ production (due to the short residence time of atomic~H on grains), and the ratio thermalizes to $\sim$$10^{-3}$. At intermediate densities, however, the ratio is boosted due to the desorption of $\rm H_2$ formed on grains. We have checked that if we initialize the simulations with an ortho/para ratio of 0.1 (across the core), as opposed to the ratio of $10^{-3}$ assumed above, we recover virtually identical results at $t = 10^6\,\rm yr$ of evolution, and with generally only a factor of $\sim$2 differences to the fiducial simulations at earlier times, for the species plotted in Fig.\,\ref{fig:abundances}.

Our predictions for the $\rm H_2$ ortho/para ratio as well as the abundance of atomic H are expected to be influenced by the assumption of $k_{\rm diss, CR}(\mbox{o$\rm H_2$}) = k_{\rm diss, CR}(\mbox{p$\rm H_2$}) = k_{\rm diss, CR}(\mbox{$\rm H_2$})$, which we had to make here due to the lack of a CR-induced photodissociation rate for ortho-$\rm H_2$. Some differences are expected especially in the outer core, or in warmer regions in general, where the relative amount of ortho-$\rm H_2$ is appreciable, though at present we cannot quantify the magnitude of the effect. The CR-induced photodissociation rate for ortho-$\rm H_2$ will be addressed in a future work.

\section{Discussion and conclusions} \label{s.Conclusion}

Here, we have tested two different values for the CR-induced $\rm {H_2}$ photodissociation rate coefficient $k_{\rm diss, CR}({\rm H_2})$: i) our new estimate of $0.831 \zeta$, and ii) $24.7 \zeta$ which was previously obtained from data tables on the Leiden database. The previously tabulated value derived from the general calculation of the CR-induced photorates for all 115 species included in the database.
The final cross section in the database has the intensity contributions of o$\rm H_2$ and p$\rm H_2$ scaled to o:p = 3:1. Thus, the derived value of $24.7 \zeta$ cannot be deemed appropriate for the CR-induced dissociation of $\rm H_2$ under molecular cloud conditions. For the present paper, we have recalculated $k_{\rm diss, CR}({\rm H_2})$ assuming $\rm H_2$ o:p = 0:1 and using the p$\rm H_2$ cross section, yielding $k_{\rm diss, CR}({\rm H_2})$ = $0.831 \zeta$.

By running a series of chemical simulations probing typical physical conditions across a molecular cloud, we showed that the value of $k_{\rm diss, CR}({\rm H_2})$ is central for a realistic estimation of the abundances of molecules like methanol whose formation on grains relies on hydrogenation processes. As a specific example, using an unrealistically high CR-induced $\rm H_2$ photodissociation rate in a chemical model will result in an overestimate of the methanol abundance in the gas phase and on grains -- in the present case the predicted methanol abundances were spread over approximately two orders of magnitude among the simulations that we ran. Interestingly, the new photodissociation rate derived here (0.831$\zeta$) has only a limited (approximately factor-of-two level) effect on predicted abundances, as compared to models that do not include the CR-induced $\rm H_2$ photodissociation process.

Finally, we comment on the very high value of $k_{\rm diss, CR}({\rm H}_2) = 404 \zeta$ derived by P24. This value was calculated by integrating the p$\rm H_2$ emission spectrum over the $\rm H_2$ photodissociation cross section given in the Leiden database. However, at the time of calculation it was not known that the cross section contains the implicit assumption of o:p=3:1, and hence the final result is erroneous. For this reason, we do not show in the present paper results for chemical simulations run using this value for the dissociation rate. We stress that all of the other (that is, for species other than $\rm H_2$) new predictions on ionization and dissociation rates presented in P24 are based on internally consistent calculations and are recommended for adoption in chemical models.

To conclude, we strongly recommend the adoption of our newly predicted value $k_{\rm diss, CR}({\rm H_2})$ = $0.831 \zeta$ for CR-induced photodissociation in chemical models. The previous value of $24.7\zeta$ should not be used, and in fact it has now been removed from the Leiden database.

\acknowledgments
O.S., P.C., T.G., and S.J. acknowledge the financial support of the Max Planck Society. M.P. acknowledges the INAF grant 2023 MERCATOR (``MultiwavelEngth signatuRes of Cosmic rAys in sTar-fOrming Regions'') and 
the INAF grant 2024 ENERGIA (``ExploriNg low-Energy cosmic Rays throuGh theoretical InvestigAtions at INAF''). D.G. acknowledges support by INAF grant PACIFISM (``PArtiCles, Ionization and Fields in the InterStellar Medium''). H.R.H. acknowledges support from PEPR Origins (``X MARKS the SPOT''). E.R.  acknowledges the support of part of this work  by the Thematic Action ``Physique et Chimie du Milieu Interstellaire'' (PCMI) of INSU Programme National ``Astro'', with contributions from CNRS Physique \& CNRS Chimie, CEA, and CNES. All authors thank the referees for their constructive comments that were very helpful for improving the manuscript.

\bibliographystyle{aasjournal}
\bibliography{photodiss_refs.bib}

@ARTICLE{Sipila19b,
       author = {{Sipil{\"a}}, O. and {Caselli}, P. and {Harju}, J.},
        title = "{Modeling deuterium chemistry in starless cores: full scrambling versus proton hop}",
      journal = {\aap},
         year = 2019,
        month = nov,
       volume = {631},
          eid = {A63},
        pages = {A63},
          doi = {10.1051/0004-6361/201936416},
archivePrefix = {arXiv},
       eprint = {1909.04341},
 primaryClass = {astro-ph.GA},
       adsurl = {https://ui.adsabs.harvard.edu/abs/2019A&A...631A..63S}
}

@ARTICLE{Caselli22,
       author = {{Caselli}, Paola and {Pineda}, Jaime E. and {Sipil{\"a}}, Olli and {Zhao}, Bo and {Redaelli}, Elena and {Spezzano}, Silvia and {Maureira}, Maria Jos{\'e} and {Alves}, Felipe and {Bizzocchi}, Luca and {Bourke}, Tyler L. and {Chac{\'o}n-Tanarro}, Ana and {Friesen}, Rachel and {Galli}, Daniele and {Harju}, Jorma and {Jim{\'e}nez-Serra}, Izaskun and {Keto}, Eric and {Li}, Zhi-Yun and {Padovani}, Marco and {Schmiedeke}, Anika and {Tafalla}, Mario and {Vastel}, Charlotte},
        title = "{The Central 1000 au of a Prestellar Core Revealed with ALMA. II. Almost Complete Freeze-out}",
      journal = {\apj},
         year = 2022,
        month = apr,
       volume = {929},
       number = {1},
          eid = {13},
        pages = {13},
          doi = {10.3847/1538-4357/ac5913},
archivePrefix = {arXiv},
       eprint = {2202.13374},
 primaryClass = {astro-ph.SR},
       adsurl = {https://ui.adsabs.harvard.edu/abs/2022ApJ...929...13C}
}

@ARTICLE{Sipila19a,
       author = {{Sipil{\"a}}, O. and {Caselli}, P. and {Redaelli}, E. and {Juvela}, M. and
         {Bizzocchi}, L.},
        title = "{Why does ammonia not freeze out in the centre of pre-stellar cores?}",
      journal = {\mnras},
         year = "2019",
        month = "Jul",
       volume = {487},
       number = {1},
        pages = {1269-1282},
          doi = {10.1093/mnras/stz1344},
archivePrefix = {arXiv},
       eprint = {1905.02384},
 primaryClass = {astro-ph.GA},
       adsurl = {https://ui.adsabs.harvard.edu/abs/2019MNRAS.487.1269S}
}

@ARTICLE{Harju25,
       author = {{Harju}, J. and {Caselli}, P. and {Sipil{\"a}}, O. and {Spezzano}, S. and {Belloche}, A. and {Bizzocchi}, L. and {Pineda}, J.~E. and {Redaelli}, E. and {Wyrowski}, F.},
        title = "{Statistical nuclear spin ratios of deuterated ammonia in the pre-stellar core L1544}",
      journal = {\aap},
         year = 2025,
        month = aug,
       volume = {700},
          eid = {A141},
        pages = {A141},
          doi = {10.1051/0004-6361/202555336},
archivePrefix = {arXiv},
       eprint = {2507.04825},
 primaryClass = {astro-ph.GA},
       adsurl = {https://ui.adsabs.harvard.edu/abs/2025A&A...700A.141H}
}

@ARTICLE{Harju24,
       author = {{Harju}, J. and {Pineda}, J.~E. and {Sipil{\"a}}, O. and {Caselli}, P. and {Belloche}, A. and {Wyrowski}, F. and {Riedel}, W. and {Redaelli}, E. and {Vasyunin}, A.~I.},
        title = "{Nuclear spin ratios of deuterated ammonia in prestellar cores. LAsMA observations of H-MM1 and Oph D}",
      journal = {\aap},
         year = 2024,
        month = feb,
       volume = {682},
          eid = {A8},
        pages = {A8},
          doi = {10.1051/0004-6361/202346578},
archivePrefix = {arXiv},
       eprint = {2311.08006},
 primaryClass = {astro-ph.GA},
       adsurl = {https://ui.adsabs.harvard.edu/abs/2024A&A...682A...8H}
}

@ARTICLE{JimenezRedondo25,
       author = {{Jim{\'e}nez-Redondo}, Miguel and {Sipil{\"a}}, Olli and {Dahl}, Robin and {Caselli}, Paola and {Jusko}, Pavol},
        title = "{Cl+ and HCl+ in Reaction with $\rm H_2$ and Isotopologues: A Glance into H Abstraction and Indirect Exchange at Astrophysical Conditions}",
      journal = {ACS Earth and Space Chemistry},
         year = 2025,
        month = mar,
       volume = {9},
       number = {3},
        pages = {782-788},
          doi = {10.1021/acsearthspacechem.4c00414},
archivePrefix = {arXiv},
       eprint = {2502.10022},
 primaryClass = {astro-ph.EP},
       adsurl = {https://ui.adsabs.harvard.edu/abs/2025ESC.....9..782J}
}

@ARTICLE{Heays17,
   author = {{Heays}, A.~N. and {Bosman}, A.~D. and {van Dishoeck}, E.~F.
    },
    title = "{Photodissociation and photoionisation of atoms and molecules of astrophysical interest}",
  journal = {\aap},
archivePrefix = "arXiv",
   eprint = {1701.04459},
 primaryClass = "astro-ph.SR",
     year = 2017,
    month = jun,
   volume = 602,
      eid = {A105},
    pages = {A105},
      doi = {10.1051/0004-6361/201628742},
   adsurl = {http://adsabs.harvard.edu/abs/2017A%26A...602A.105H}
}

@ARTICLE{Padovani24,
       author = {{Padovani}, Marco and {Galli}, Daniele and {Scarlett}, Liam H. and {Grassi}, Tommaso and {Rehill}, Una S. and {Zammit}, Mark C. and {Bray}, Igor and {Fursa}, Dmitry V.},
        title = "{Ultraviolet H$_{2}$ luminescence in molecular clouds induced by cosmic rays}",
      journal = {\aap},
         year = 2024,
        month = feb,
       volume = {682},
          eid = {A131},
        pages = {A131},
          doi = {10.1051/0004-6361/202348168},
archivePrefix = {arXiv},
       eprint = {2312.02062},
 primaryClass = {astro-ph.GA},
       adsurl = {https://ui.adsabs.harvard.edu/abs/2024A&A...682A.131P}
}

@ARTICLE{Keto15,
   author = {{Keto}, E. and {Caselli}, P. and {Rawlings}, J.},
    title = "{The dynamics of collapsing cores and star formation}",
  journal = {\mnras},
archivePrefix = "arXiv",
   eprint = {1410.5889},
 primaryClass = "astro-ph.SR",
     year = 2015,
    month = feb,
   volume = 446,
    pages = {3731-3740},
      doi = {10.1093/mnras/stu2247},
   adsurl = {http://adsabs.harvard.edu/abs/2015MNRAS.446.3731K}
}

@ARTICLE{Padovani18a,
       author = {{Padovani}, Marco and {Ivlev}, Alexei V. and {Galli}, Daniele and {Caselli}, Paola},
        title = "{Cosmic-ray ionisation in circumstellar discs}",
      journal = {\aap},
         year = 2018,
        month = jun,
       volume = {614},
          eid = {A111},
        pages = {A111},
          doi = {10.1051/0004-6361/201732202},
archivePrefix = {arXiv},
       eprint = {1803.09348},
 primaryClass = {astro-ph.HE},
       adsurl = {https://ui.adsabs.harvard.edu/abs/2018A&A...614A.111P}
}

@ARTICLE{Sipila15a,
   author = {{Sipil{\"a}}, O. and {Caselli}, P. and {Harju}, J.},
    title = "{Benchmarking spin-state chemistry in starless core models}",
  journal = {\aap},
archivePrefix = "arXiv",
   eprint = {1501.04825},
     year = 2015,
    month = jun,
   volume = 578,
      eid = {A55},
    pages = {A55},
      doi = {10.1051/0004-6361/201424364},
   adsurl = {http://adsabs.harvard.edu/abs/2015A%26A...578A..55S}
}

@ARTICLE{Wakelam15,
   author = {{Wakelam}, V. and {Loison}, J.-C. and {Herbst}, E. and {Pavone}, B. and 
	{Bergeat}, A. and {B{\'e}roff}, K. and {Chabot}, M. and {Faure}, A. and 
	{Galli}, D. and {Geppert}, W.~D. and {Gerlich}, D. and {Gratier}, P. and 
	{Harada}, N. and {Hickson}, K.~M. and {Honvault}, P. and {Klippenstein}, S.~J. and 
	{Le Picard}, S.~D. and {Nyman}, G. and {Ruaud}, M. and {Schlemmer}, S. and 
	{Sims}, I.~R. and {Talbi}, D. and {Tennyson}, J. and {Wester}, R.
	},
    title = "{The 2014 KIDA Network for Interstellar Chemistry}",
  journal = {\apjs},
archivePrefix = "arXiv",
   eprint = {1503.01594},
     year = 2015,
    month = apr,
   volume = 217,
      eid = {20},
    pages = {20},
      doi = {10.1088/0067-0049/217/2/20},
   adsurl = {http://adsabs.harvard.edu/abs/2015ApJS..217...20W}
}

@ARTICLE{Sipila24,
       author = {{Sipil{\"a}}, O. and {Caselli}, P. and {Juvela}, M.},
        title = "{Impact of ice growth on the physical and chemical properties of dense cloud cores: I. Monodisperse grains}",
      journal = {\aap},
         year = 2024,
        month = oct,
       volume = {690},
          eid = {A280},
        pages = {A280},
          doi = {10.1051/0004-6361/202450824},
archivePrefix = {arXiv},
       eprint = {2408.17202},
 primaryClass = {astro-ph.GA},
       adsurl = {https://ui.adsabs.harvard.edu/abs/2024A&A...690A.280S}
}

@ARTICLE{Lupi21,
       author = {{Lupi}, Alessandro and {Bovino}, Stefano and {Grassi}, Tommaso},
        title = "{On the low ortho-to-para H$_{2}$ ratio in star-forming filaments}",
      journal = {\aap},
         year = 2021,
        month = oct,
       volume = {654},
          eid = {L6},
        pages = {L6},
          doi = {10.1051/0004-6361/202142145},
archivePrefix = {arXiv},
       eprint = {2109.02655},
 primaryClass = {astro-ph.GA},
       adsurl = {https://ui.adsabs.harvard.edu/abs/2021A&A...654L...6L}
}

@ARTICLE{Herbst73,
   author = {{Herbst}, E. and {Klemperer}, W.},
    title = "{The Formation and Depletion of Molecules in Dense Interstellar Clouds}",
  journal = {\apj},
     year = 1973,
    month = oct,
   volume = 185,
    pages = {505-534},
      doi = {10.1086/152436},
   adsurl = {http://adsabs.harvard.edu/abs/1973ApJ...185..505H}
}

@ARTICLE{Hasegawa93a,
   author = {{Hasegawa}, T.~I. and {Herbst}, E.},
    title = "{New gas-grain chemical models of quiescent dense interstellar clouds - The effects of H2 tunnelling reactions and cosmic ray induced desorption}",
  journal = {\mnras},
     year = 1993,
    month = mar,
   volume = 261,
    pages = {83-102},
   adsurl = {http://adsabs.harvard.edu/abs/1993MNRAS.261...83H}
}

@ARTICLE{Prasad83,
   author = {{Prasad}, S.~S. and {Tarafdar}, S.~P.},
    title = "{UV radiation field inside dense clouds - Its possible existence and chemical implications}",
  journal = {\apj},
     year = 1983,
    month = apr,
   volume = 267,
    pages = {603-609},
      doi = {10.1086/160896},
   adsurl = {http://adsabs.harvard.edu/abs/1983ApJ...267..603P}
}

@ARTICLE{Sternberg87,
   author = {{Sternberg}, A. and {Dalgarno}, A. and {Lepp}, S.},
    title = "{Cosmic-ray-induced photodestruction of interstellar molecules in dense clouds}",
  journal = {\apj},
     year = 1987,
    month = sep,
   volume = 320,
    pages = {676-682},
      doi = {10.1086/165585},
   adsurl = {http://adsabs.harvard.edu/abs/1987ApJ...320..676S}
}

@ARTICLE{Gredel89,
   author = {{Gredel}, R. and {Lepp}, S. and {Dalgarno}, A. and {Herbst}, E.
	},
    title = "{Cosmic-ray-induced photodissociation and photoionization rates of interstellar molecules}",
  journal = {\apj},
     year = 1989,
    month = dec,
   volume = 347,
    pages = {289-293},
      doi = {10.1086/168117},
   adsurl = {http://adsabs.harvard.edu/abs/1989ApJ...347..289G}
}

@ARTICLE{Garrod06,
       author = {{Garrod}, R.~T. and {Herbst}, E.},
        title = "{Formation of methyl formate and other organic species in the warm-up phase of hot molecular cores}",
      journal = {\aap},
         year = 2006,
        month = oct,
       volume = {457},
       number = {3},
        pages = {927-936},
          doi = {10.1051/0004-6361:20065560},
archivePrefix = {arXiv},
       eprint = {astro-ph/0607560},
 primaryClass = {astro-ph},
       adsurl = {https://ui.adsabs.harvard.edu/abs/2006A&A...457..927G}
}

@ARTICLE{Ruffle01,
       author = {{Ruffle}, Deborah P. and {Herbst}, Eric},
        title = "{New models of interstellar gas-grain chemistry - II. Surface photochemistry in quiescent cores}",
      journal = {\mnras},
         year = 2001,
        month = apr,
       volume = {322},
       number = {4},
        pages = {770-778},
          doi = {10.1046/j.1365-8711.2001.04178.x},
       adsurl = {https://ui.adsabs.harvard.edu/abs/2001MNRAS.322..770R}
}

@ARTICLE{Gredel87,
       author = {{Gredel}, R. and {Lepp}, S. and {Dalgarno}, A.},
        title = "{The C/CO Ratio in Dense Interstellar Clouds}",
      journal = {\apjl},
         year = 1987,
        month = dec,
       volume = {323},
        pages = {L137},
          doi = {10.1086/185073},
       adsurl = {https://ui.adsabs.harvard.edu/abs/1987ApJ...323L.137G}
}

@ARTICLE{Hrodmarsson23,
       author = {{Hrodmarsson}, H.~R. and {van Dishoeck}, E.~F.},
        title = "{Photodissociation and photoionization of molecules of astronomical interest. Updates to the Leiden photodissociation and photoionization cross section database}",
      journal = {\aap},
         year = 2023,
        month = jul,
       volume = {675},
          eid = {A25},
        pages = {A25},
          doi = {10.1051/0004-6361/202346645},
       adsurl = {https://ui.adsabs.harvard.edu/abs/2023A&A...675A..25H}
}

@ARTICLE{Padovani09,
       author = {{Padovani}, M. and {Galli}, D. and {Glassgold}, A.~E.},
        title = "{Cosmic-ray ionization of molecular clouds}",
      journal = {\aap},
         year = 2009,
        month = jul,
       volume = {501},
       number = {2},
        pages = {619-631},
          doi = {10.1051/0004-6361/200911794},
archivePrefix = {arXiv},
       eprint = {0904.4149},
 primaryClass = {astro-ph.SR},
       adsurl = {https://ui.adsabs.harvard.edu/abs/2009A&A...501..619P}
}

@ARTICLE{Padovani22,
       author = {{Padovani}, Marco and {Bialy}, Shmuel and {Galli}, Daniele and {Ivlev}, Alexei V. and {Grassi}, Tommaso and {Scarlett}, Liam H. and {Rehill}, Una S. and {Zammit}, Mark C. and {Fursa}, Dmitry V. and {Bray}, Igor},
        title = "{Cosmic rays in molecular clouds probed by H$_{2}$ rovibrational lines. Perspectives for the James Webb Space Telescope}",
      journal = {\aap},
         year = 2022,
        month = feb,
       volume = {658},
          eid = {A189},
        pages = {A189},
          doi = {10.1051/0004-6361/202142560},
archivePrefix = {arXiv},
       eprint = {2201.08457},
 primaryClass = {astro-ph.GA},
       adsurl = {https://ui.adsabs.harvard.edu/abs/2022A&A...658A.189P}
}

@ARTICLE{Ivlev21,
       author = {{Ivlev}, Alexei V. and {Silsbee}, Kedron and {Padovani}, Marco and {Galli}, Daniele},
        title = "{Rigorous Theory for Secondary Cosmic-Ray Ionization}",
      journal = {\apj},
         year = 2021,
        month = mar,
       volume = {909},
       number = {2},
          eid = {107},
        pages = {107},
          doi = {10.3847/1538-4357/abdc27},
archivePrefix = {arXiv},
       eprint = {2101.05803},
 primaryClass = {astro-ph.GA},
       adsurl = {https://ui.adsabs.harvard.edu/abs/2021ApJ...909..107I}
}

@ARTICLE{Scarlett23,
       author = {{Scarlett}, Liam H. and {Rehill}, Una S. and {Zammit}, Mark C. and {Mori}, Nicolas A. and {Bray}, Igor and {Fursa}, Dmitry V.},
        title = "{Elastic scattering and rotational excitation of H$_{2}$ by electron impact: Convergent close-coupling calculations}",
      journal = {\pra},
         year = 2023,
        month = jun,
       volume = {107},
       number = {6},
          eid = {062804},
        pages = {062804},
          doi = {10.1103/PhysRevA.107.062804},
       adsurl = {https://ui.adsabs.harvard.edu/abs/2023PhRvA.107f2804S}
}

@ARTICLE{Abgrall93a,
       author = {{Abgrall}, H. and {Roueff}, E. and {Launay}, F. and {Roncin}, J. -Y. and {Subtil}, J. -L.},
        title = "{Table of Lyman band system of molecular hydrogen.}",
      journal = {\aaps},
         year = 1993,
        month = oct,
       volume = {101},
        pages = {273-321},
       adsurl = {https://ui.adsabs.harvard.edu/abs/1993A&AS..101..273A}
}

@ARTICLE{Abgrall93b,
       author = {{Abgrall}, H. and {Roueff}, E. and {Launay}, F. and {Roncin}, J. -Y. and {Subtil}, J. -L.},
        title = "{Table of the Werner band system of molecular hydrogen.}",
      journal = {\aaps},
         year = 1993,
        month = oct,
       volume = {101},
        pages = {323-362},
       adsurl = {https://ui.adsabs.harvard.edu/abs/1993A&AS..101..323A}
}

@ARTICLE{Abgrall00,
       author = {{Abgrall}, H. and {Roueff}, E. and {Drira}, I.},
        title = "{Total transition probability and spontaneous radiative dissociation of B, C, B' and D states of molecular hydrogen}",
      journal = {\aaps},
         year = 2000,
        month = jan,
       volume = {141},
        pages = {297-300},
          doi = {10.1051/aas:2000121},
       adsurl = {https://ui.adsabs.harvard.edu/abs/2000A&AS..141..297A}
}

@ARTICLE{Abgrall97,
       author = {{Abgrall}, H. and {Roueff}, E. and {Liu}, Xianming and {Shemansky}, D.~E.},
        title = "{The Emission Continuum of Electron-excited Molecular Hydrogen}",
      journal = {\apj},
         year = 1997,
        month = may,
       volume = {481},
       number = {1},
        pages = {557-566},
          doi = {10.1086/304017},
       adsurl = {https://ui.adsabs.harvard.edu/abs/1997ApJ...481..557A}
}

@ARTICLE{Roueff19,
       author = {{Roueff}, E. and {Abgrall}, H. and {Czachorowski}, P. and {Pachucki}, K. and {Puchalski}, M. and {Komasa}, J.},
        title = "{The full infrared spectrum of molecular hydrogen}",
      journal = {\aap},
         year = 2019,
        month = oct,
       volume = {630},
          eid = {A58},
        pages = {A58},
          doi = {10.1051/0004-6361/201936249},
archivePrefix = {arXiv},
       eprint = {1909.11585},
 primaryClass = {physics.atom-ph},
       adsurl = {https://ui.adsabs.harvard.edu/abs/2019A&A...630A..58R}
}

@ARTICLE{Furuya16,
       author = {{Furuya}, K. and {van Dishoeck}, E.~F. and {Aikawa}, Y.},
        title = "{Reconstructing the history of water ice formation from HDO/H$_{2}$O and D$_{2}$O/HDO ratios in protostellar cores}",
      journal = {\aap},
         year = 2016,
        month = feb,
       volume = {586},
          eid = {A127},
        pages = {A127},
          doi = {10.1051/0004-6361/201527579},
archivePrefix = {arXiv},
       eprint = {1512.04291},
 primaryClass = {astro-ph.GA},
       adsurl = {https://ui.adsabs.harvard.edu/abs/2016A&A...586A.127F}
}

@ARTICLE{Abgrall93c,
       author = {{Abgrall}, H. and {Roueff}, E. and {Launay}, F. and {Roncin}, J.~Y. and {Subtil}, J.~L.},
        title = "{The Lyman and Werner Band Systems of Molecular Hydrogen}",
      journal = {Journal of Molecular Spectroscopy},
         year = 1993,
        month = feb,
       volume = {157},
       number = {2},
        pages = {512-523},
          doi = {10.1006/jmsp.1993.1040},
       adsurl = {https://ui.adsabs.harvard.edu/abs/1993JMoSp.157..512A}
}

@ARTICLE{Glass-Maujean2013,
       author = {{Glass-Maujean}, M. and {Jungen}, Ch. and {Spielfiedel}, A. and {Schmoranzer}, H. and {Tulin}, I. and {Knie}, A. and {Reiss}, P. and {Ehresmann}, A.},
        title = "{Experimental and theoretical studies of excited states of H$_{2}$ observed in the absorption spectrum: I.}",
      journal = {Journal of Molecular Spectroscopy},
         year = 2013,
        month = nov,
       volume = {293},
        pages = {1-10},
          doi = {10.1016/j.jms.2013.09.010},
       adsurl = {https://ui.adsabs.harvard.edu/abs/2013JMoSp.293....1G}
}

@ARTICLE{Glass-Maujean2012a,
       author = {{Glass-Maujean}, M. and {Schmoranzer}, H. and {Haar}, I. and {Knie}, A. and {Reiss}, P. and {Ehresmann}, A.},
        title = "{The J = 1 para levels of the v = 0 to 6 np singlet Rydberg series of molecular hydrogen revisited}",
      journal = {\jcp},
         year = 2012,
        month = apr,
       volume = {136},
       number = {13},
        pages = {134301-134301},
          doi = {10.1063/1.3697967},
       adsurl = {https://ui.adsabs.harvard.edu/abs/2012JChPh.136m4301G}
}

@ARTICLE{Glass-Maujean2012b,
       author = {{Glass-Maujean}, M. and {Schmoranzer}, H. and {Haar}, I. and {Knie}, A. and {Reiss}, P. and {Ehresmann}, A.},
        title = "{The J = 2 ortho levels of the v = 0 to 6 np singlet Rydberg series of molecular hydrogen revisited}",
      journal = {\jcp},
         year = 2012,
        month = aug,
       volume = {137},
       number = {8},
          eid = {084303},
        pages = {084303},
          doi = {10.1063/1.4742311},
       adsurl = {https://ui.adsabs.harvard.edu/abs/2012JChPh.137h4303G}
}

@ARTICLE{Walmsley2004,
       author = {{Walmsley}, C.~M. and {Flower}, D.~R. and {Pineau des For{\^e}ts}, G.},
        title = "{Complete depletion in prestellar cores}",
      journal = {\aap},
         year = 2004,
        month = may,
       volume = {418},
        pages = {1035-1043},
          doi = {10.1051/0004-6361:20035718},
archivePrefix = {arXiv},
       eprint = {astro-ph/0402493},
 primaryClass = {astro-ph},
       adsurl = {https://ui.adsabs.harvard.edu/abs/2004A&A...418.1035W}
}

@ARTICLE{Padovani18b,
       author = {{Padovani}, Marco and {Galli}, Daniele and {Ivlev}, Alexei V. and {Caselli}, Paola and {Ferrara}, Andrea},
        title = "{Production of atomic hydrogen by cosmic rays in dark clouds}",
      journal = {\aap},
         year = 2018,
        month = nov,
       volume = {619},
          eid = {A144},
        pages = {A144},
          doi = {10.1051/0004-6361/201834008},
archivePrefix = {arXiv},
       eprint = {1809.04168},
 primaryClass = {astro-ph.GA},
       adsurl = {https://ui.adsabs.harvard.edu/abs/2018A&A...619A.144P}
}

\end{document}